\documentclass[twocolumn,showpacs,showkeys,preprintnumbers]{revtex4}
\usepackage{amssymb}
\usepackage{amsfonts}
\usepackage{amsmath}
\usepackage{graphicx}
\usepackage{dcolumn}
\usepackage{bm}
\usepackage{epsfig}

\begin{document}

\title{Polytropic representation of the kinetic pressure tensor of non-ideal
magnetized fluids in equilibrium toroidal structures}
\author{Claudio Cremaschini$^{a}$, Ji\v{r}\'{\i} Kov\'{a}\v{r}$^{a}$, Zden%
\v{e}k Stuchl\'{\i}k$^{a}$ and Massimo Tessarotto$^{b,a}$}
\affiliation{$^{a}$Research Centre for Theoretical Physics and Astrophysics, Institute of
Physics, Silesian University in Opava, Bezru\v{c}ovo n\'{a}m.13, CZ-74601
Opava, Czech Republic}
\affiliation{$^{b}$Department of Mathematics and Geosciences, University of Trieste, Via
Valerio 12, 34127 Trieste, Italy}
\date{\today }

\begin{abstract}
Non-ideal fluids are generally subject to the occurrence of non-isotropic
pressure tensors, whose determination is fundamental in order to
characterize their dynamical and thermodynamical properties. This requires
the implementation of theoretical frameworks provided by appropriate
microscopic and statistical kinetic approaches in terms of which continuum
fluid fields are obtained. In this paper the case of non-relativistic
magnetized fluids forming equilibrium toroidal structures in external
gravitational fields is considered. Analytical solutions for the kinetic
distribution function are explicitly constructed, to be represented by a
Chapman-Enskog expansion around a Maxwellian equilibrium. In this way,
different physical mechanisms responsible for the generation of
non-isotropic pressures are identified and proved to be associated with the
kinetic constraints imposed on single and collective particle dynamics by
phase-space symmetries and magnetic field. As a major outcome, the validity
of a polytropic representation for the kinetic pressure tensors
corresponding to each source of anisotropy is established, whereby
directional pressures exhibit a specific power-law functional dependence on
fluid density. The astrophysical relevance of the solution for the
understanding of fluid plasma properties in accretion-disc environments is
discussed.
\end{abstract}

\keywords{Non-ideal Magnetized Fluid; Equation Of State; Polytropic
Pressure;\ \ Pressure Anisotropy; Non-isotropic Pressure Tensor. \\
\underline{Corresponding Author:} Claudio Cremaschini - email:
claudiocremaschini@gmail.com}
\maketitle

\section{Introduction}

In fluid dynamics, in the absence of an independent evolution equation for
the fluid scalar pressure $P$, the prescription of the equation of state
(EoS) for $P$ - or closure condition (CC) for the relevant fluid equations -
appears generally a complex problem of crucial importance. This involves
representing the same fluid pressure in terms of a suitable set of
independent fluid fields (i.e., the fluid state) and external parameters
describing the physical, dynamical and thermodynamical properties of the
fluid system under consideration. The EoS, besides carrying information
about the physical state of the system \cite{Kaur}, may permit - in turn -
also the analytical and/or numerical\ treatment for the relevant fluid
equation \cite{Caprioli2,Sironi1,Trauble,Karas1}. In many applications the
EoS is expressed in terms of a single isotropic scalar pressure $P$
prescribed by an equation of state of the type $P=P\left( n,T,\mathbf{V}%
\right) $. This may depend both on the local fluid number density $n$, the
local scalar temperature $T$ and also the local fluid velocity field $V$.
Its precise form should in principle be determined separately based on
phenomenological models and/or microscopic (i.e., kinetic) physics that
pertains the structure and interactions occurring among the same
constituents of the fluid, e.g., atoms, charges or molecules, possibly
subject to external fields \cite{Eyal,Poor,Muir}. A particular case that
belongs to this category pertains to ideal fluids.\ As such they are
intended here as continuum systems assumed to be described at microscopic
level by a phase-space statistics determined by a local (and possibly
Maxwellian) kinetic distribution function (KDF). For ideal systems of this
type the pressure $P\ $is a position-dependent scalar function expressed by
the well-known ideal relationship (in IS dimensional units)$\ P=nT$ \cite{Li}%
.

An alternative model route often pursued in hydrodynamics consists in
treating the scalar pressure $P$ as a function of the fluid mass density $%
\rho $ only \cite{Tejeda1,Rhodes,Caprioli1,Ovalle}. This yields the
so-called polytropic form of the EoS, which in customary notation is written
as%
\begin{equation}
P=\kappa \rho ^{\Gamma },  \label{poly-1}
\end{equation}%
where $\kappa $ is a suitable dimensional numerical factor of
proportionality (frequently taken to be constant), while $\Gamma $ is the
polytropic exponent factor (or polytropic index). The choice of $\kappa $
and $\Gamma $ distinguishes the physical settings and the kind of relevant
physical effects retained by the polytropic EoS, including for example
isentropic and isothermal processes or the so-called stiff EoS implemented
for relativistic fluids \cite{Tejeda2,Tejeda3}. Remarkably, the power-law
functional form of Eq.(\ref{poly-1}) relates directly the pressure to the
density and does not depend on other state variables like the temperature.
For this reason, the polytropic form of EoS finds a wide range of
applicability that spans both relativistic and non-relativistic gases \cite%
{Boccelli}, degenerate matter and astrophysical fluids and plasmas \cite%
{coll-2014,coll-2016,Liva,universe2020}.

However, actual physical fluid systems may be expected to deviate in several
ways from the ideal-fluid state \cite{Boon}. The same concept of having a
single scalar pressure, eventually assigning to the polytropic relation of
type (\ref{poly-1}) a sort of universal character, may represent a
restrictive assumption, an incomplete characterization of the fluid system
or even a conceptually-wrong statement. This can be particularly relevant in
the case of collisionless - or even suitably weakly collisional -
astrophysical neutral or charged fluids (respectively fluids or plasmas)
and/or magnetized plasmas. In fact, in these systems, the action of
gravitational and electromagnetic fields in combination with other effects
like radiation fields, conservation laws, dissipation or trapping phenomena,
boundary-layer conditions and geometrical configuration-space constraints
can lead to the onset of so-called non-ideal fluids \cite%
{APJS,Asenjo1,Zhu,Tolman}. The non-ideal features pertain the definition of
the EoS. More precisely, in the present framework a fluid is said to be
non-ideal if its EoS ceases to be represented by a scalar pressure and
becomes instead expressed in terms of a pressure tensor exhibiting
directional pressures, and therefore yielding a pressure (and temperature)
anisotropy \cite{Hall,Holm,Douanla}. Hence, the non-ideal case amounts to
replacing the pressure $P$ with a pressure tensor of the type%
\begin{equation}
P\rightarrow \underline{\underline{\Pi }}\equiv \Pi _{ab},
\label{pressure tensor}
\end{equation}%
where, for non-relativistic treatments, indices $a,b$ range from $1$ to $3$.

The proper understanding of the physical effects contributing to the
generation of a pressure tensor and the correct determination of its
mathematical structure demand the adoption of statistical treatments in
terms of appropriate kinetic theories. It is only within the framework of a
kinetic approach that is possible to gain a comprehensive description able
to deal with microscopic field interactions as well as single-particle or
collective system dynamics. The issue is particularly relevant in
collisionless or weakly-collisional systems \cite{Ott} (see also definition
below), composed of either neutral or charged non-degenerate matter, to be
described within the framework of Vlasov theory. According to such a kinetic
description, the fundamental quantity is represented by the species KDF $%
f_{s}$, where the subscript "$s$" identifies the species index. For
non-relativistic regimes, the KDF is defined on the single-particle
6th-dimensional phase-space and its dynamics is determined by the Vlasov
equation. In Lagrangian form, the latter is written as%
\begin{equation}
\frac{d}{dt}f_{s}\left( \mathbf{x}\left( t\right) ,t\right) =0,
\label{vlasov-1}
\end{equation}%
where\ in general the function $f_{\mathrm{s}}$\ can still depend explicitly
on the time $t$, while here $\mathbf{x}\equiv \left( \mathbf{r},\mathbf{v}%
\right) $ denotes the particle state. The velocity integrals of the KDF
define appropriate fluid fields, namely physical observables, while the
velocity moments of the kinetic equation (\ref{vlasov-1}) determine the
corresponding set of continuum fluid equations. It must be stressed that, in
the framework of the kinetic approach, the prescription of the closure
conditions (e.g., the pressure tensor) for the fluid equations becomes
unique and self-consistent once a solution of the Vlasov equation is
obtained. In particular, in collisionless systems the kinetic effects must
be properly retained because the corresponding fluid closure conditions can
become non-trivial \cite{Cr2011}. This originates from the fact that
collisionless systems can develop phase-space anisotropies, both at
equilibrium as well as in non-stationary configurations, including
instability or turbulence phenomena, which are carried by the KDF through
its functional dependences \cite{Mika,Kunz}.

In fact, from the kinetic point of view, the occurrence of a pressure tensor
(\ref{pressure tensor}) corresponds to a continuum fluid described by a
non-isotropic, namely non-Maxwellian, KDF. Deviations from a Maxwellian
function must be treated in statistical way as they generally amount to the
onset of phase-space anisotropies that ultimately show up in the fluid
domain. For example, the latter can be responsible for the occurrence of
collective drift-velocities, like those associated with diamagnetic, finite
Larmor-radius (FLR) and energy-correction effects, shear-flow phenomena as
well as magnetic field generation (kinetic dynamo)\ by local current
densities in plasmas \cite{Cr2013b,Cr2013c}. Similarly, temperature and
pressure anisotropies as well as non-vanishing heat fluxes can arise on this
basis \cite{Egedal}. Phenomena of this kind are expected to be relevant in
particular for equilibrium or slowly-time varying configurations, namely
evolving on time-scales longer than other characteristic time-scales of the
system \cite{Soni}. As such they can occur in both relativistic and
non-relativistic regimes for neutral gravitating matter and charged or
magnetized fluids \cite{Bek2,Carter1,Carter2}, in systems subject to
geometric symmetry constraints or in absence of spatial symmetries \cite%
{Cr2013,PoP2014-2} and even in the presence of electromagnetic radiation
fields \cite{Bek1,PoP2014-1}. For this reason, due to the intrinsic
complexity of the task caused by the underlying non-linear statistical
dynamics, obtaining an explicit and possibly analytical representation of
the KDF corresponding to a non-ideal fluid may result a formidable task.

To focus on the issue in greater detail, in the following we restrict our
attention to non-relativistic collisionless magnetized plasmas at
equilibrium in toroidal structures which are subject to the presence of
external gravitational and electromagnetic (EM) fields. Concrete
realizations of systems of this kind include in particular the case of
plasmas belonging to accretion-disk or hot coronal environments in the
surrounding of compact objects, for which the characteristic axial symmetry
can represent a reasonable approximation. Thus, given a cylindrical
coordinate reference system $\left( R,\varphi ,z\right) $, the coordinate $%
\varphi $ is ignorable and identifies the direction of symmetry. The sources
of anisotropy carried by the KDF that are considered in the present research
are due to the existence of microscopic conservation laws generating the
consequent appearance of phase-space and velocity-space constraints on the
same form of the KDF. These kinematic/dynamical constraints affect the
velocity dispersion relations defining the statistical directional plasma
temperatures in terms of weighted integrals over the equilibrium KDF, which
are then translated into anisotropy of the pressure tensor. More precisely,
the following two distinct possible mechanisms are treated:

\textit{Case A - }Temperature anisotropy induced by conservation of particle
canonical momentum imposed by configuration-space assumption of axisymmetry.
The anisotropy arises between the azimuthal spatial direction $\varphi $ and
the vertical and radial directions $\left( R,z\right) $. We refer to this
case as the tangential pressure anisotropy.

\textit{Case B} - Temperature anisotropy generated by conservation of
particle magnetic moment. This represents intrinsically a velocity-space
symmetry constraint associated with the Larmor gyration motion of charges
around magnetic field lines. We refer this case as the magnetic-moment
pressure anisotropy. The resulting spatial direction of anisotropy depends
in this case on the orientation of the magnetic field with respect to the
coordinate system $\left( R,\varphi ,z\right) $. To exemplify the physical
mechanism and make possible an analytical comparison with the tangential
anisotropy case, we treat here the magnetic field configuration
corresponding to the purely-vertical topology. This case is relevant in
astrophysical studies of toroidal accretion-disc systems.

Given these premises, the goals of the present paper are summarized as
follows:

1)\ First, we proceed introducing the theoretical formalism required for the
construction of equilibrium KDFs appropriate for the treatment of the
anisotropic effects A and B listed above.

2)\ Second, by identifying suitable kinetic regimes, the same equilibrium
solutions are then represented by a Chapman-Enskog expansion around a
Maxwellian equilibrium distribution in terms of a suitable dimensionless
parameter $\varepsilon \ll 1$ (see definition below). In this way the two
physical mechanisms A and B, responsible for the generation of non-isotropic
pressures, can be singled-out and their role inspected analytically. This
permits to treat the non-ideal corrections to the EoS in a perturbative way,
by establishing corresponding representations for the tensor pressure of the
type%
\begin{equation}
\underline{\underline{\Pi }}=P\underline{\underline{I}}+\varepsilon 
\underline{\underline{\pi }}+O\left( \varepsilon ^{2}\right) .  \label{1-A}
\end{equation}%
Here, $P$ is the leading-order scalar pressure carried by the Maxwellian
KDF, $\underline{\underline{I}}$ is the identity matrix and $\underline{%
\underline{\pi }}$ is the $O\left( \varepsilon \right) $ term of pressure
tensor carrying the whole physical informations that characterize the
non-ideal fluid contributions to the EoS due to either effects A and B.

3) Third, to implement an iterative scheme that permits to represent the
tensor $\underline{\underline{\Pi }}$ in Eq.(\ref{1-A}) in polytropic form.
Namely, in the sense that each non-vanishing entry of the tensor $\underline{%
\underline{\Pi }}$ can be finally expressed as a precise function of
power-law mass density $\rho $ with a characteristic polytropic index $%
\Gamma $. This establishes the sought polytropic representation of the
pressure tensor for non-ideal fluids. The result is fundamental also because
it allows to unveil the physical implications of the kinetic anisotropies as
they reflect on the single directional pressures of the fluid, together with
their functional dependence on the mass-density spatial profile.

4) Fourth, to compare the form of the polytropic representation of the
tensor $\underline{\underline{\pi }}$ for the two cases A and B. This
involves also the corresponding differential contributions arising in the
momentum equation, which mark the difference of the non-ideal fluid
configuration from the ideal-fluid solution. The target of this analysis
consists in the possibility of pointing out the characteristic fingerprints
of each effect A and B on physical observables of real system configurations.

5)\ The final point concerns the qualitative investigation of the possible
astrophysical relevance of the solution obtained here for the understanding
of non-ideal plasma properties in accretion-disc environments. This includes
a comparison with alternative theoretical models based simply on the
ideal-fluid assumption of having a continuum medium with isotropic
temperature and scalar pressure. More precisely, we address here the problem
of the observational estimate of temperature profiles of magnetized plasma
systems and the consequent interpolation of their energy content and mass
density distribution in the presence of the aforementioned non-ideal kinetic
effects of kind A and B.

\section{General physical assumptions}

In this section we introduce the basic physical assumptions required for the
development of the subsequent theory. These concern the physical state of
the plasma system, configuration-space geometry and phase-space conservation
laws as well as the electromagnetic and gravitational fields representations.

We consider the case of a magnetized plasma subject to an
externally-generated gravitational field together with the possible
occurrence of electrostatic interactions. Possible weakly-dissipative
effects (Coulomb collisions and turbulence) and EM radiation-reaction
effects are ignored. We shall assume that the KDF and the EM fields
associated with the plasma obey the system of Vlasov-Maxwell equations, with
Maxwell's equations being considered in the quasi-static approximation. For
definiteness, we shall consider here a plasma consisting of $s$-species of
charged particles which are characterized by proper mass $M_{s}$ and total
charge $Z_{s}e$. The generic KDF $f_{s}=f_{s}\left( \mathbf{x}\left(
t\right) ,t\right) =f_{s}\left( \mathbf{r},\mathbf{v},t\right) $ is defined
in the phase-space $\Gamma =\Gamma _{\mathbf{r}}\times \Gamma _{\mathbf{v}}$%
, with $\Gamma _{\mathbf{r}}$ and $\Gamma _{\mathbf{v}}$ being the
configuration and velocity spaces respectively, and its dynamical evolution
is prescribed by the Vlasov equation (\ref{vlasov-1}).

In this treatment the plasma is taken to be: a) non-relativistic, in the
sense that it has non--relativistic particles and species flow velocities,
that the gravitational field can be treated within the classical Newtonian
theory, and that the non-relativistic Vlasov kinetic equation is used as the
dynamical equation for the KDF;\ b)\ axisymmetric, so that the relevant
dynamical variables characterizing the plasma (e.g., the fluid fields) are
independent of the azimuthal angle $\varphi $, when referred to a set of
cylindrical coordinates $\mathbf{r}=(R,\varphi ,z)$. In addition, a kinetic
treatment appropriate for the description of collisionless fluids is
developed. In the present context, the latter are identified by the
requirement that the species mean free path of plasma particles, denoted
with $\lambda _{mfp}$, is much greater than the largest characteristic scale
length of the plasma $L_{scale}$. Thus, collisionless fluids are such that $%
\lambda _{mfp}\gg L_{scale}$. We refer instead to the weakly-collisional
fluid regime if the inequality $\lambda _{mfp}\gtrsim L_{scale}$ applies. In
practice, the collisionless assumption can be verified a posteriori once the
kinetic solution is known by explicit evaluation of the species scale-length 
$\lambda _{mfp}$. The latter must take into consideration the rate of
occurrence of microscopic collisions among fluid particles, possibly in
combination with the simultaneous action of confining mechanisms associated
with magnetic fields (e.g., the Larmor rotation).

On general grounds, quasi-stationary configurations are treated, namely
characterized by solutions which are slowly-varying in time. This condition
is also referred to as equilibrium configuration. For a generic physical
quantity $G$ which depends on spatial coordinates $\mathbf{r}$ and time $t$,
the quasi-stationarity is expressed by letting in the following $G=G\left( 
\mathbf{r},\varepsilon ^{k}t\right) $, with $\varepsilon \ll 1$ being a
small dimensionless parameter to be suitably defined (see below) and $k\geq
1 $ being an integer.

We focus on solutions for the equilibrium magnetic field $\mathbf{B}$ which
admit, at least locally, a family of nested axisymmetric toroidal magnetic
surfaces $\left\{ \psi (\mathbf{r},\varepsilon ^{k}t)\right\} \equiv \left\{
\psi (\mathbf{r},\varepsilon ^{k}t)=const.\right\} $, where $\psi $ denotes
the poloidal magnetic flux of $\mathbf{B}$. We then require the EM field to
be slowly-varying in time, i.e., of the form 
\begin{equation}
\left[ \mathbf{E}(\mathbf{r},\varepsilon ^{k}t),\mathbf{B}(\mathbf{r}%
,\varepsilon ^{k}t)\right] .  \label{b0}
\end{equation}%
In particular, we assume the magnetic field to be represented as%
\begin{equation}
\left. \mathbf{B}\equiv \nabla \times \mathbf{A}=\mathbf{B}^{self}(\mathbf{r}%
,\varepsilon ^{k}t)+\mathbf{B}^{ext}(\mathbf{r},\varepsilon ^{k}t),\right.
\label{b1}
\end{equation}%
where $\mathbf{B}^{self}$ and $\mathbf{B}^{ext}$ denote the self-generated
magnetic field produced by the toroidal plasma and a finite external
axisymmetric magnetic field (vacuum field). In the most general
configuration, both contributions can exhibit non-vanishing azimuthal and
poloidal components, to be represented as%
\begin{eqnarray}
\mathbf{B}^{ext} &=&I_{ext}(\mathbf{r},\varepsilon ^{k}t)\nabla \varphi
+\nabla \psi _{ext}(\mathbf{r},\varepsilon ^{k}t)\times \nabla \varphi , \\
\mathbf{B}^{self} &=&I_{self}(\mathbf{r},\varepsilon ^{k}t)\nabla \varphi
+\nabla \psi _{self}(\mathbf{r},\varepsilon ^{k}t)\times \nabla \varphi .
\label{bself}
\end{eqnarray}%
Hence, the total magnetic field takes the form%
\begin{equation}
\mathbf{B}=I(\mathbf{r},\varepsilon ^{k}t)\nabla \varphi +\nabla \psi (%
\mathbf{r},\varepsilon ^{k}t)\times \nabla \varphi ,  \label{B FIELD}
\end{equation}%
where $\mathbf{B}_{T}\equiv I(\mathbf{r},\varepsilon ^{k}t)\nabla \varphi $
and $\mathbf{B}_{P}\equiv \nabla \psi (\mathbf{r},\varepsilon ^{k}t)\times
\nabla \varphi $ are the corresponding toroidal (i.e., azimuthal) and
poloidal components respectively, with $I\equiv I_{ext}+I_{self}$ and $\psi
\equiv \psi _{ext}+\psi _{self}$. For the case of interest here it is
assumed that the magnetic field is primarily externally generated, so that $%
\mathbf{B}^{self}\ll \mathbf{B}^{ext}$.

Finally, charged particles are assumed to be subject to the effective
potential $\Phi _{s}^{eff}(\mathbf{r},\varepsilon ^{k}t)$ defined as%
\begin{equation}
\Phi _{s}^{eff}(\mathbf{r},\varepsilon ^{k}t)\equiv \Phi (\mathbf{r}%
,\varepsilon ^{k}t)+\frac{M_{s}}{Z_{s}e}\Phi _{G}(\mathbf{r},\varepsilon
^{k}t),  \label{efp}
\end{equation}%
with $\Phi (\mathbf{r},\varepsilon ^{k}t)$ and $\Phi _{G}(\mathbf{r}%
,\varepsilon ^{k}t)$ denoting the electrostatic potential generated by the
plasma charge density and the gravitational potential. In the following we
shall neglect the contribution of the plasma to $\Phi _{G}$, which can
therefore be assumed as being stationary and to be externally-generated,
e.g., by the central compact object (like black-hole systems) in the case of
accretion discs. However, in the non-relativistic regime, i.e., for
Newtonian gravity, the potential $\Phi _{G}$ can be conveniently identified
uniquely with a spherically-symmetric (and therefore also axially-symmetric)
potential associated with the central mass of the compact object.
Nevertheless, we stress that the formalism proposed here is general enough
to allow modelling in principle more complex gravitational potentials, which
can also take into account the contribution of the mass distribution of the
fluid disc itself \cite{Carter2}. In any case, it is important to recall
that the action of the gravitational field marks a substantial distinctive
point of difference between laboratory and astrophysical plasmas and fluids.
The gravitational interaction affects in a non-trivial way the
single-particle dynamics by means of mutual balance among gravitational,
electrostatic and magnetic forces. This effect is then inherited by the
statistical distribution of charged particles, and therefore it shows up in
the kinetic formalism. As shown below, the main contributions of the
gravitational potential are through the definition of particle invariants
and the resulting form of fluid number density.

Concerning single-particle dynamics and conservation laws, we have the
following set of invariants:

1)\ The particle canonical momentum $p_{\varphi s}$ conjugate to the
azimuthal angle $\varphi $:%
\begin{equation}
p_{\varphi s}=M_{s}R\mathbf{v}\cdot \mathbf{e}_{\varphi }+\frac{Z_{s}e}{c}%
\psi (\mathbf{r},\varepsilon ^{k}t)\equiv \frac{Z_{s}e}{c}\psi _{\ast s}.
\label{pifi}
\end{equation}

2)\ The particle total energy $E_{s}$:%
\begin{equation}
E_{s}=\frac{M_{s}}{2}v^{2}\mathbf{+}{Z_{s}e}\Phi _{s}^{eff}(\mathbf{r}%
,\varepsilon ^{k}t)\equiv Z_{s}e\Phi _{\ast s}.  \label{energia}
\end{equation}

3) The particle magnetic moment $m_{s}^{\prime }\equiv \frac{Z_{s}e}{M_{s}c}%
p_{\phi ^{\prime }s}^{\prime }$, which is by definition an adiabatic
invariant, where according to gyrokinetic theory $p_{\phi ^{\prime
}s}^{\prime }=\partial \mathcal{L}_{s}^{\prime }/\partial \overset{\cdot }{%
\phi ^{\prime }}$ is the momentum conjugate to the gyrophase angle of Larmor
rotation around magnetic field lines.

We can now define the dimensionless parameter $\varepsilon \equiv \min
\left\{ \varepsilon _{s}\right\} $, where the species parameter $\varepsilon
_{s}$ is used in the introduction of appropriate plasma orderings required
for the subsequent analytical determination of EoS. Notice that, although in
principle several characteristic dimensionless parameters can be identified
in plasma systems and be used for the definition of corresponding kinetic
regimes (see for example Ref.\cite{Cr2012}), in the present case we
formulate the subsequent theory in terms of the unique parameter $%
\varepsilon $. In fact, without loss of generality, this permits to gain a
simpler formal representation of the theory and to enhance focus on the
problem of the polytropic representation of the EoS. In detail, $\varepsilon
_{s}$ is prescribed in such a way to be independent of single-particle
velocity but to be related to the characteristic species perpendicular and
parallel thermal velocities (defined with respect to the local magnetic
field direction). They are defined respectively by $v_{\perp ths}=\left\{
T_{\perp s}/M_{s}\right\} ^{1/2}$\ and $v_{\parallel ths}=\left\{
T_{\parallel s}/M_{s}\right\} ^{1/2}$, with $T_{\perp s}$\ and $T_{\parallel
s}$\ denoting here the species perpendicular and parallel temperatures.
Thus, $\varepsilon _{s}$ is defined as $\varepsilon _{s}\equiv \frac{r_{Ls}}{%
L}$, where $r_{Ls}=v_{\perp ths}/\Omega _{cs}$\ is the species average
Larmor radius with cyclotron frequency $\Omega _{cs}$, while $L$\ is the
minimum scale-length characterizing the spatial variations of all of the
fluid fields associated with the KDF and of the EM fields. The parameter $%
\varepsilon _{s}$\ is customarily adopted in gyrokinetic theory for the
analytical asymptotic representation of the magnetic moment. In fact, it is
possible to determine $\mathcal{L}_{s}^{\prime }$ so that $m_{s}^{\prime }$
is an adiabatic invariant of arbitrary order in $\varepsilon _{s}\ll 1$. In
particular, the leading-order approximation is $m_{s}^{\prime }\simeq \mu
_{s}^{\prime }\equiv \frac{M_{s}w^{\prime 2}}{2B^{\prime }}$. In addition
here we also ignore higher-order correction terms originating from inverse
gyrokinetic transformation, regarded of $O\left( \varepsilon _{s}^{k}\right) 
$, with $k\geq 1$, thus assuming that%
\begin{equation}
m_{s}^{\prime }\simeq \mu _{s}^{\prime }=\mu _{s}\left[ 1+O\left(
\varepsilon _{s}^{k}\right) \right] ,  \label{ordering-mm2}
\end{equation}%
where%
\begin{equation}
\mu _{s}\equiv \frac{M_{s}w^{2}}{2B},  \label{mm-2}
\end{equation}%
and $w$ denotes the magnitude of particle velocity perpendicular to the
magnetic field line, while $B$ is the local magnitude of magnetic field.

Finally, it is necessary to introduce additional ordering assumptions on the
relative magnitudes of the terms defining the conserved canonical momentum (%
\ref{pifi}) and energy (\ref{energia}). This is required in order to afford
subsequently an analytical treatment of the kinetic and fluid solutions, to
be detailed in the following sections. Again, as is customary in plasma
physics, it is convenient to prescribe these orderings in terms of
characteristic species thermal velocities. In fact, for nearly-Maxwellian
kinetic solutions, like those considered in the present study (see below),
this provides the appropriate average estimates that characterize observable
features of the fluid. As anticipated above, this analysis is carried out
here in terms of the single small dimensionless parameter $\varepsilon _{s}$%
, although in principle several distinctive parameters can be introduced
(see Ref.\cite{Cr2012}).

The first ordering concerns the canonical momentum and is expressed by
requiring that%
\begin{equation}
\left\vert \frac{M_{s}Rv_{ths}}{\frac{Z_{s}e}{c}\psi }\right\vert \sim
O\left( \varepsilon _{s}^{j}\right) ,  \label{a1}
\end{equation}%
where $v_{ths}\equiv \sup \left\{ v_{\parallel ths},v_{\perp ths}\right\} $\
and here $j\geq 1$. The previous quantity effectively measures the ratio
between the toroidal angular momentum $L_{\varphi s}\equiv M_{s}Rv_{\varphi
} $\ and the magnetic contribution to the toroidal canonical momentum, for
all particles in which $v_{\varphi }$\ is of the order of $v_{ths}$, while $%
\psi $\ is assumed as being non-vanishing. Similarly, the second ordering
concerns the particle energy, imposing that 
\begin{equation}
\left\vert \frac{\frac{M_{s}}{2}v_{ths}^{2}}{{Z_{s}e}\Phi _{s}^{eff}}%
\right\vert \sim O\left( \varepsilon _{s}^{j}\right) ,  \label{a2}
\end{equation}%
where $j\geq 1$. This quantity measures the ratio between particle kinetic
and potential energies, for all thermal particles, with $\Phi _{s}^{eff}$\
being assumed as non-vanishing. As a consequence of order equivalences (\ref%
{a1}) and (\ref{a2}), the following asymptotic expansions apply:%
\begin{eqnarray}
\psi _{\ast s} &=&\psi \left[ 1+O\left( \varepsilon _{s}^{j}\right) \right] ,
\label{a3} \\
\Phi _{\ast s} &\mathbf{=}&\Phi _{s}^{eff}\left[ 1+O\left( \varepsilon
_{s}^{j}\right) \right] .  \label{a4}
\end{eqnarray}

The choice of the exponents $k$\ in (\ref{ordering-mm2}) and $j$\ in (\ref%
{a1}) and (\ref{a2}) then defines the kind of kinetic regime. We stress that
this in turn corresponds to effectively select the kind of physical
properties of the system to be retained in the analytical solution, and the
relative importance of each corresponding conservation law. For example, for 
$k=j=1$\ one recovers for the collisionless plasma the strongly-magnetized
and strong effective potential energy regime treated in Ref.\cite{Cr2012}
where contributions arising from the previous expansions are regarded to be
of first-order in\ $\varepsilon _{s}$. The values of $k$\ and $j$\
appropriate for the target of the present study are discussed in Section IV
below.

\section{Kinetic equilibrium}

In this section we develop the formalism for the representation of
analytical solutions of the KDF corresponding to kinetic equilibria
describing non-ideal fluid configurations. As shown below, these type of
kinetic solutions are found to be intrinsically different from the
Maxwellian function. In order to set the treatment on general grounds, we
adopt a general formalism that can be later applied to investigate the two
cases A and B mentioned above. Hence, the characterization of the general
kinetic solutions obtained here to the different physical scenarios of
interest is left to subsequent specialized sections.

The starting point is the identification of the set $I_{\ast s}$ of
invariants common to each configuration. Given the underlying validity of
stationarity and axisymmetry assumptions, these are necessarily identified
with energy and canonical momentum, so that we can write%
\begin{equation}
I_{\ast s}=\left( E_{s},\psi _{\ast s}\right) .
\end{equation}%
Hence, inclusion of the set $I_{\ast s}$ is essential to all configurations
treated below, so that $E_{s}$ and $\psi _{\ast s}$ acquire an ubiquitous
character. The method of invariants for the construction of analytical
solutions of the Vlasov equation describing kinetic equilibria in
collisionless plasmas is implemented (see Ref.\cite{Cr2011}). This amounts
to expressing the equilibrium KDF in terms of the set of invariants and
appropriate kinetic constraints that warrant its consistency with the Vlasov
equation. Accordingly, the general form of the species equilibrium (i.e.,
quasi-stationary)\ KDF is expressed as%
\begin{equation}
f_{\ast s}=f_{\ast s}\left( I_{\ast s},Y_{\ast s},\Lambda _{\ast
s},\varepsilon ^{k}t\right) ,  \label{form}
\end{equation}%
where $Y_{\ast s}$ carries the source of anisotropy, while $\Lambda _{\ast
s} $ denotes the so-called structure functions \cite{Cr2011}, i.e.,\
suitably-defined functions which depend implicitly on the particle state $%
\left( \mathbf{r},\mathbf{v}\right) $. In order for $f_{\ast s}$ to be an
adiabatic invariant, both $Y_{\ast s}$ and $\Lambda _{\ast s}$ must also be
functions of the adiabatic invariants. This restriction is referred to as a
kinetic constraint. For definiteness, both $f_{\ast s}$ and $\Lambda _{\ast
s}$ are assumed to be analytic functions. We notice that the dependence on
particle energy generates the Maxwellian character of the solution, while
that on the canonical momentum permits inclusion of toroidal fluid velocity
(shifted Maxwellian function). In contrast, the dependence on $Y_{\ast s}$
expresses the non-ideal character of the solution, associated with
phase-space anisotropies. The particular realization of $Y_{\ast s}$ makes
possible the treatment of the different scenarios A and B and must be
specified according to each physical context of interest.

To determine an explicit representation for $f_{\ast s}$ according to Eq.(%
\ref{form}), we introduce the following additional requirements:

1)\ The KDF $f_{\ast s}$ must be a strictly-positive real function and it
must be summable, in the sense that the velocity moments on the
velocity-space $\Gamma _{\mathbf{v}}$ of the form%
\begin{equation}
\Xi _{s}(\mathbf{r},\varepsilon ^{k}t)=\int_{\Gamma _{\mathbf{v}%
}}d^{3}vK_{s}(\mathbf{r},\mathbf{v},\varepsilon ^{k}t)f_{\ast s}
\label{velomo}
\end{equation}%
must exist for a suitable ensemble of weight functions $\left\{ K_{s}(%
\mathbf{x},\varepsilon ^{k}t)\right\} $, to be prescribed in terms of
polynomials of arbitrary degree defined with respect to components of the
velocity vector field $\mathbf{v}$. This can be warranted by imposing the
minimum requirement that, at least, the equilibrium KDF has a Gaussian-like
dependence on particle energy, namely expressible in terms of the
characteristic exponential Maxwellian form $f_{\ast s}\sim e^{-v^{2}}$.

2) The KDF $f_{\ast s}$ must be characterized by non-uniform fluid fields,
including in particular non-uniform species-dependent number density, flow
velocity (with dominant azimuthal flow component according to the physical
characterization of accretion discs)\ and non-isotropic temperature, to be
suitably prescribed in terms of $Y_{\ast s}$ and the structure functions.

Given these prescriptions, a particular solution for the equilibrium KDF is
taken of the form%
\begin{equation}
f_{\ast s}=\frac{\eta _{\ast s}}{\left( 2\pi /M_{s}\right) ^{3/2}T_{\ast
s}^{3/2}}\exp \left\{ -\frac{H_{\ast s}}{T_{\ast s}}-\alpha _{\ast s}Y_{\ast
s}\right\} ,  \label{f*}
\end{equation}%
which we refer to as the \emph{equilibrium non-isotropic KDF}. Concerning
the notation, here%
\begin{equation}
H_{\ast s}\equiv E_{s}-\frac{Z_{s}e}{c}\psi _{\ast s}\Omega _{\ast s},
\end{equation}%
where $\Omega _{\ast s}$ is a generalized frequency to be properly related
to the fluid azimuthal rotation frequency, while $\eta _{\ast s}$ is
referred to as the generalized species pseudo-density, $T_{\ast s}$ is the
generalized species isotropic temperature and $\alpha _{\ast s}$ denotes the
anisotropy-strength function. As an alternative, invoking the definitions (%
\ref{pifi}) and (\ref{energia}), Eq.(\ref{f*}) can be equivalently written as%
\begin{equation}
f_{\ast s}=\frac{\eta _{\ast s}\exp \left[ \frac{X_{\ast s}}{T_{\ast s}}%
\right] }{\left( 2\pi /M_{s}\right) ^{3/2}T_{\ast s}^{3/2}}\exp \left\{ -%
\frac{M_{s}\left( \mathbf{v}-\mathbf{V}_{\ast s}\right) ^{2}}{2T_{\ast s}}%
-\alpha _{\ast s}Y_{\ast s}\right\} ,  \label{f*-bis}
\end{equation}%
where $\mathbf{V}_{\ast s}=\mathbf{e}_{\varphi }R\Omega _{\ast s}$ and%
\begin{equation}
X_{\ast s}\equiv M_{s}\frac{\left\vert \mathbf{V}_{\ast s}\right\vert ^{2}}{2%
}+\frac{Z_{s}e}{c}\psi \Omega _{\ast s}-Z_{s}e\Phi _{s}^{eff}.
\end{equation}

In Eq.(\ref{f*}) the structure functions are identified with the set $%
\left\{ \Lambda _{\ast s}\right\} \equiv \left\{ \eta _{\ast s},T_{\ast
s},\Omega _{\ast s},\alpha _{\ast s}\right\} $. The characterization of the
solution is finally completed by appropriate specification of the kinetic
constraints for $\left\{ \Lambda _{\ast s}\right\} $. In the most general
case, these must be assigned as follows:%
\begin{equation}
\Lambda _{\ast s}=\Lambda _{s}\left( \psi _{\ast s},\Phi _{\ast s}\right) .
\label{kincon}
\end{equation}%
Notice that at this stage the set $\left\{ \Lambda _{\ast s}\right\} $\
cannot be directly identified with particular fluid fields. This is because
of the functional form of the structure functions according to Eq.(\ref%
{kincon}), and due to the non-Maxwellian character of the solution implied
by the term $\alpha _{\ast s}Y_{\ast s}$. As a consequence, the latter fluid
fields must be consistently computed as velocity moments of the KDF from Eq.(%
\ref{velomo}).

\section{Perturbative theory and analytical solutions}

In this section we develop a perturbative theory appropriate for the
analytical treatment of the functional dependences carried by the
equilibrium KDF. In particular, the solution (\ref{f*}) is further required
to admit a Chapman-Enskog asymptotic representation of the type%
\begin{equation}
f_{\ast s}=f_{Ms}\left[ 1+\varepsilon _{s}\delta f_{s}+O\left( \varepsilon
_{s}^{2}\right) \right] ,  \label{ch-1}
\end{equation}%
where $f_{Ms}$ and $\delta f_{s}$ are respectively the leading-order and the
first-order (i.e., $O\left( \varepsilon _{s}\right) $) contributions. For
clarity, we have left the symbol $\varepsilon _{s}$ in front of $\delta
f_{s} $ to emphasize this is a linear term in the same expansion parameter.
In the present work we require the term $f_{Ms}$\ to be identified with the
customary Maxwellian distribution. As such, the Chapman-Enskog series sought
here amounts to effectively developing an asymptotic representation of the
equilibrium KDF that permits the analytical description of nearly-Maxwellian
plasmas. The deviations from the purely-Maxwellian case due to collisionless
dynamics that give rise to non-ideal fluid effects are therefore carried by
the linear $O\left( \varepsilon _{s}\right) $\ contribution $\delta f_{s}$.
This permits an unambiguous interpretation of their physical meaning, to be
determined invoking the ordering assumptions introduced in Section 2. In
this respect, the most general form of solution for $\delta f_{s}$\ should
contain both explicit and implicit functional dependences, associated
respectively with particle invariants and differential treatment of
functional form of structure functions (like in Ref.\cite{Cr2011}). Given
the intrinsic completeness of kinetic theory, the simultaneous treatment of
all these contributions would certainly generate a complex solution that
could obscure the precise target of the research. For this reason, in order
to obtain an analytical expression of $\delta f_{s}$ able to single out the
precise terms of interest that generate the types of temperature anisotropy
treated here, we must properly set the value of the asymptotic expansions.
More precisely, it is required that the first-order term $\delta f_{s}$\
carries only the explicit functional dependences associated with $Y_{\ast s}$%
. For this reason, any additional (implicit)\ functional dependence carried
by the structure functions from ordering equivalences (\ref{a1}) and (\ref%
{a2}) or due to inverse gyrokinetic transformation of magnetic moment
representation (\ref{ordering-mm2}) must be accordingly treated as being of
higher-order. This requires considering a different kind of plasma regime
with respect to those treated in Ref.\cite{Cr2012}. The latter configuration
can be met by setting hereafter $k=j=2$\ in Eqs.(\ref{ordering-mm2}) and (%
\ref{a1})-(\ref{a2}). Such a choice appears therefore physically motivated
and is admissible within the framework of construction method of exact
kinetic equilibrium solutions based on the method of invariants.

We start from the implicit functional dependences associated with the
phase-functions $\psi _{\ast s}$\ and $\Phi _{\ast s}$ contained in the
structure functions $\left\{ \Lambda _{\ast s}\right\} $ according to the
prescription (\ref{kincon}). The perturbative theory is obtained by
performing on $f_{\ast s}$ a double-Taylor expansion for $\left\{ \Lambda
_{\ast s}\right\} $ of the form%
\begin{eqnarray}
\Lambda _{\ast s} &=&\Lambda _{s}+\left( \psi _{\ast s}-\psi \right) \left[ 
\frac{\partial \Lambda _{\ast s}}{\partial \psi _{\ast s}}\right] 
_{\substack{ \psi _{\ast s}=\psi  \\ \Phi _{\ast s}=\Phi _{s}^{eff}}}  \notag
\\
&&+\left( \Phi _{\ast s}-\Phi _{s}^{eff}\right) \left[ \frac{\partial
\Lambda _{\ast s}}{\partial \Phi _{\ast s}}\right] _{\substack{ \psi _{\ast
s}=\psi  \\ \Phi _{\ast s}=\Phi _{s}^{eff}}}.  \label{espan0}
\end{eqnarray}%
The terms proportional to $\left( \psi _{\ast s}-\psi \right) $ are referred
to as diamagnetic corrections, while those proportional to $\left( \Phi
_{\ast s}-\Phi _{s}^{eff}\right) $ as energy-correction contributions. Since
these expansion terms contain the gradients of the structure functions
across magnetic and energy equipotential surfaces, they permit in principle
the consistent description of kinetic solutions characterized by non-uniform
fluid fields and their gradient behavior in equilibrium configurations. In
the following we omit to provide the explicit representation of these
gradient contributions as this goes beyond the scope of the present notes
and is not necessary for the following calculations. By invoking Eqs.(\ref%
{a1})-(\ref{a2}) and the validity of Eqs.(\ref{a3})-(\ref{a4}), due to the
motivations expressed above, letting $j=2$\ yields gradient contributions
that are of second-order in the dimensionless parameter $\varepsilon _{s}$.
Therefore, according to the ordering imposed here, the Taylor-expanded
structure functions $\left\{ \Lambda _{\ast s}\right\} $ can be written
simply as%
\begin{equation}
\Lambda _{\ast s}=\Lambda _{s}\left( \psi ,\Phi _{s}^{eff}\right) \left[
1+O\left( \varepsilon _{s}^{2}\right) \right] .  \label{exp-lambda}
\end{equation}%
When we apply this expansion to $f_{\ast s}$ we therefore obtain the
following preliminary Chapman-Enskog representation:%
\begin{equation}
f_{\ast s}=f_{s}^{\left( 0\right) }\left[ 1+O\left( \varepsilon
_{s}^{2}\right) \right] ,  \label{solo}
\end{equation}%
where $f_{s}^{\left( 0\right) }$ denotes the leading-order KDF, while the $%
O\left( \varepsilon _{s}^{2}\right) $ corrections arise from the
perturbative treatment of the structure functions. More precisely, from Eq.(%
\ref{f*-bis}) it follows that $f_{s}^{\left( 0\right) }$ is of the form%
\begin{equation}
f_{s}^{\left( 0\right) }=\frac{\eta _{s}\exp \left[ \frac{X_{s}}{T_{s}}%
\right] }{\left( 2\pi /M_{s}\right) ^{3/2}T_{s}^{3/2}}\exp \left\{ -\frac{%
M_{s}\left( \mathbf{v}-\mathbf{V}_{s}\right) ^{2}}{2T_{s}}-\alpha
_{s}Y_{\ast s}\right\} ,  \label{f0-a}
\end{equation}%
which identifies the \emph{equilibrium non-isotropic Gaussian KDF}, where%
\textbf{\ }$\mathbf{V}_{s}=\Omega _{s}R\mathbf{e}_{\varphi }$, while the
function $X_{s}$ is%
\begin{equation}
X_{s}\equiv M_{s}\frac{R^{2}\Omega _{s}^{2}}{2}+\frac{Z_{s}e}{c}\psi \Omega
_{s}^{\left( 0\right) }-Z_{s}e\Phi _{s}^{eff}.
\end{equation}

The final step to reach the representation (\ref{ch-1}) concerns the
appropriate treatment of the anisotropy contribution $\alpha _{s}Y_{\ast s}$
in the exponential function of Eq.(\ref{f0-a}). In order to single out the
non-isotropic features associated with $\alpha _{s}Y_{\ast s}$\ with respect
to the others non-Maxwellian corrections generated by the structure
functions that are of $O\left( \varepsilon _{s}^{2}\right) $, the same $%
\alpha _{s}Y_{\ast s}$\ contribution must be small compared to the
leading-order Maxwellian term, but stronger than the diamagnetic and
energy-correction terms. Accordingly, this must represent the only surviving
linear contribution of the Chapman-Enskog series due to explicit particle
invariants with non-isotropic velocity dependence. Therefore, we must
require that for thermal particles with thermal velocity $v_{th}$ the
following additional ordering applies:%
\begin{equation}
\left. \alpha _{s}Y_{\ast s}\right\vert _{v_{\varphi \sim v_{th}}}\sim
O\left( \varepsilon _{s}\right) .  \label{ordera}
\end{equation}%
As a consequence, we can finally write the equilibrium KDF in the desired
form%
\begin{equation}
f_{s}^{\left( 0\right) }=f_{Ms}\left[ 1+\varepsilon _{s}\delta f_{s}\right] ,
\end{equation}%
where respectively%
\begin{equation}
f_{Ms}\equiv \frac{\eta _{s}\exp \left[ \frac{X_{s}}{T_{s}}\right] }{\left(
2\pi /M_{s}\right) ^{3/2}T_{s}^{3/2}}\exp \left\{ -\frac{M_{s}\left( \mathbf{%
v}-\mathbf{V}_{s}\right) ^{2}}{2T_{s}}\right\} ,  \label{max-0}
\end{equation}%
is the isotropic Maxwellian distribution, while $\delta f_{s}\sim O\left(
\varepsilon _{s}\right) $ is given by%
\begin{equation}
\delta f_{s}=-\alpha _{s}Y_{\ast s}.
\end{equation}

We can now give a precise physical meaning to the leading-order structure
functions $\Lambda _{s}\left( \psi ,\Phi _{s}^{eff}\right) $. In detail, the
quantity $\eta _{s}\exp \left[ \frac{X_{s}}{T_{s}}\right] $ is related to
the plasma number density, $T_{s}$ represents the Maxwellian isotropic
temperature, $\mathbf{V}_{s}$ is the azimuthal flow velocity and finally $%
\alpha _{s}$ measures the strength of the anisotropy effect. The
representation (\ref{ch-1})\ is therefore particularly convenient to isolate
the source of explicit phase-space anisotropy expressed by $Y_{\ast s}$ with
respect to possible non-isotropic features induced by diamagnetic and
energy-correction contributions. The explicit form of $Y_{\ast s}$ must be
prescribed according to the physical scenarios A and B that we intend to
study here. However, in principle the theoretical framework established
above and the corresponding mathematical treatment possess a general
character that makes the theory susceptible of additional similar studies,
even for more complex configurations. We notice in addition that the
advantage of having a leading-order Maxwellian function lies in the
possibility of concrete application of the theory to predictions of physical
state associated with either collisionless plasmas or collisional
nearly-Maxwellian plasmas, namely systems that are allowed to exhibit weak
deviations from the purely Maxwellian collisional case, to occur on
time-scales shorter that the characteristic collisional time-scale of the
system. In both cases the theory permits to relate the features of the
temperature and pressure anisotropies to the quantity $\alpha _{s}$, which
is a physical observable (i.e., a continuum fluid field)\ that must
necessarily enter the definition of the non-ideal contributions to the
pressure tensor. Accordingly, we refer to $\alpha _{s}=\alpha _{s}\left(
\psi ,\Phi _{s}^{eff}\right) $ as the anisotropy-strength function. Finally,
the same Maxwellian character of the leading-order KDF obtained here is also
instrumental for the subsequent proof of validity of the polytropic
representation of the EoS for the non-ideal plasma described by the
equilibrium KDF $f_{\ast s}$ given by Eq.(\ref{f*-bis}).

We now proceed addressing in detail separately the two cases A and B of
interest, providing the explicit analytical solutions for the corresponding
perturbative representations of the equilibrium KDFs.

\subsection{Analytical solution: case A}

In case A we consider temperature anisotropy induced by conservation of
particle canonical momentum holding for axisymmetric systems. The analytical
perturbative solution for the equilibrium KDF is obtained by letting%
\begin{equation}
\delta f_{s}=-\alpha _{s}\psi _{\ast s}^{2}.
\end{equation}%
Explicitly, this term is represented as%
\begin{eqnarray}
\delta f_{s} &=&-\alpha _{s}\left( \frac{M_{s}cR}{Z_{s}e}v_{\varphi }+\psi
\right) ^{2}  \notag \\
&=&-\alpha _{s}\left[ \left( \frac{M_{s}cR}{Z_{s}e}\right) ^{2}v_{\varphi
}^{2}+\psi ^{2}+2\frac{M_{s}cR}{Z_{s}e}\psi v_{\varphi }\right] ,
\label{df-a}
\end{eqnarray}%
where $v_{\varphi }\equiv \mathbf{v}\cdot \mathbf{e}_{\varphi }$. As a final
result, we see that in such a configuration and under validity of previous
ordering assumptions, the species equilibrium KDF is taken to depend only on
the two invariants $\psi _{\ast s}$ and $\Phi _{\ast s}$. In particular, the
anisotropy is associated with the square dependence on particle canonical
momentum, which will be shown to correctly reproduce temperature anisotropy
arising between the azimuthal and the $\left( R,z\right) $ directions. This
kind of effect is expected to be relevant in toroidal structures where
azimuthal rotation can generate tangential pressure profiles decoupled from
vertical and radial ones.

\subsection{Analytical solution: case B}

Case B corresponds to the treatment of temperature anisotropy generated by
conservation of particle magnetic moment $m_{s}^{\prime }$. The analytical
perturbative solution for the equilibrium KDF requires identifying $\delta
f_{s}=-\alpha _{s}m_{s}^{\prime }$. On the other hand, the joint validity of
the orderings (\ref{ordering-mm2}) and (\ref{ordera}) allows to approximate
this expression for later convenience of calculation. Thus, neglecting
corrections of $O\left( \varepsilon _{s}^{k}\right) $, with $k\geq 2$, we
can require%
\begin{equation}
\delta f_{s}=-\alpha _{s}\mu _{s},
\end{equation}%
where $\mu _{s}$ is defined by Eq.(\ref{mm-2}). For the case of interest
here, we first assume that the magnetic field is primarily externally
generated so that (see also Section 2) $\mathbf{B}^{self}\ll \mathbf{B}%
^{ext} $. Furthermore, we assume that also the following relative ordering
applies:%
\begin{equation}
\mathbf{B}_{T}\ll \mathbf{B}_{P},
\end{equation}%
so that the poloidal field is dominating in the region occupied by the
plasma. Finally, in the same domain the magnetic field is taken to be purely
vertical and uniform, namely in cylindrical coordinates $(R,\varphi ,z)$%
\begin{equation}
\mathbf{B}^{ext}=(0,0,B_{z}\equiv const.).  \label{bext-1}
\end{equation}%
This component corresponds to a generating flux function $\psi $ of the type 
$\psi =\psi _{o}R$, which therefore coincides exactly with the radial
coordinate, a part from the dimensional multiplicative constant factor.
Alternatively, one can also envisage a more general solution for the
magnetic field, which however can still be approximated as being
purely-vertical in the region of the torus occupied by the fluid. This
assumption is appropriate for example in the case of so-called thin
accretion discs. The difference now is that the vertical component becomes
spatial-dependent, and one can replace the expression (\ref{bext-1}) with%
\begin{equation}
\mathbf{B}^{ext}=(0,0,B_{z}\equiv B_{z}\left( R,z\right) ).  \label{bext-2}
\end{equation}%
As a result of such topology for the magnetic field, we have that the square
velocity component in the plane orthogonal to magnetic field lines is
expressed as $w^{2}=v_{R}^{2}+v_{\varphi }^{2}$, where $v_{R}\equiv \mathbf{v%
}\cdot \mathbf{e}_{R}$ and $v_{\varphi }\equiv \mathbf{v}\cdot \mathbf{e}%
_{\varphi }$. Inserting this expression in the definition of $\mu _{s}$
yields the representation of $\delta f_{s}$ in the following desired form: 
\begin{equation}
\delta f_{s}=-\alpha _{s}\frac{M_{s}w^{2}}{2B}=-\alpha _{s}\frac{M_{s}}{2B}%
\left( v_{R}^{2}+v_{\varphi }^{2}\right) .  \label{df-b}
\end{equation}

\section{Pressure tensor}

Having obtained the analytical solution for the equilibrium KDF, we can now
proceed evaluating the corresponding fluid moment relevant for the present
investigation, to be identified with the species pressure tensor. The latter
is defined as 
\begin{equation}
\underline{\underline{\mathbf{\Pi }}}_{s}\equiv \int d\mathbf{v}M_{s}\left( 
\mathbf{v}-\mathbf{V}_{s}\right) \left( \mathbf{v}-\mathbf{V}_{s}\right)
f_{\ast s}.
\end{equation}%
In view of the analytical Chapman-Enskog representation of $f_{\ast s}$
obtained above in Eq.(\ref{ch-1}), correct through $O\left( \varepsilon
_{s}^{2}\right) $ the same tensor takes the form%
\begin{equation}
\underline{\underline{\mathbf{\Pi }}}_{s}=\int d\mathbf{v}M_{s}\left( 
\mathbf{v}-\mathbf{V}_{s}\right) \left( \mathbf{v}-\mathbf{V}_{s}\right)
f_{Ms}\left[ 1+\delta f_{s}\right] ,  \notag
\end{equation}%
where it is intended that $\delta f_{s}\sim O\left( \varepsilon _{s}\right) $
and for simplicity of notation we now omit to write explicitly the symbol $%
\varepsilon _{s}$ before it. It is easily seen that the tensor is composed
of two contributions, namely the leading-order term associated with the
Maxwellian function $f_{Ms}$ and the perturbative correction due to $\delta
f_{s}$, which expresses the non-ideal character of the fluid solution.
Hence, we can equally write%
\begin{equation}
\underline{\underline{\mathbf{\Pi }}}_{s}=\pi _{Ms}+\delta \pi _{s},
\end{equation}%
where the meaning of the symbols is understood.

We start by computing the representation due to the Maxwellian contribution $%
\pi _{Ms}$. First, we notice that the number density associated with $f_{Ms}$
and defined as%
\begin{equation}
n_{Ms}\equiv \int d\mathbf{v}f_{Ms},
\end{equation}%
is immediately computed and gives%
\begin{equation}
n_{Ms}\equiv \eta _{s}\exp \left[ \frac{X_{s}}{T_{s}}\right] .
\end{equation}%
The equilibrium Maxwellian KDF $f_{Ms}$ introduced in Eq.(\ref{max-0})\ can
therefore be written in the more familiar form as%
\begin{equation}
f_{Ms}\equiv \frac{n_{M}}{\left( 2\pi /M_{s}\right) ^{3/2}T_{s}^{3/2}}\exp
\left\{ -\frac{M_{s}\left( \mathbf{v}-\mathbf{V}_{s}\right) ^{2}}{2T_{s}}%
\right\} .
\end{equation}%
The corresponding pressure tensor is found to be isotropic in the scalar
pressure $P_{s}$ and to retain the customary representation in terms of
density and temperature given by%
\begin{equation}
\pi _{Ms}=P_{s}\underline{\underline{\mathbf{I}}}=P_{s}\left( 
\begin{array}{ccc}
1 & 0 & 0 \\ 
0 & 1 & 0 \\ 
0 & 0 & 1%
\end{array}%
\right) ,
\end{equation}%
where%
\begin{equation}
P_{s}\equiv n_{Ms}T_{s}.
\end{equation}

The non-isotropic component $\delta \pi _{s}$ is defined as%
\begin{equation}
\delta \pi _{s}=\int d\mathbf{v}M_{s}\left( \mathbf{v}-\mathbf{V}_{s}\right)
\left( \mathbf{v}-\mathbf{V}_{s}\right) f_{Ms}\delta f_{s}.  \label{deltap-0}
\end{equation}%
We can now obtain analytical expressions for $\delta \pi _{s}$ in the two
physical realizations treated here.

\subsection{Pressure anisotropy: case A}

In the case A the representation of $\delta f_{s}$ to be used in Eq.(\ref%
{deltap-0}) is provided by Eq.(\ref{df-a}). To evaluate the integrals
explicitly we first introduce the change of variables letting $\mathbf{u=v}-%
\mathbf{V}_{s}$, where $d\mathbf{u=}d\mathbf{v}$, so that expressing the
Maxwellian KDF we can write explicitly%
\begin{equation}
\delta \pi _{s}=\frac{n_{M}M_{s}}{\left( 2\pi /M_{s}\right) ^{3/2}T_{s}^{3/2}%
}\int d^{3}u\mathbf{uu}e^{-\frac{M_{s}\left( \mathbf{u}\right) ^{2}}{2T_{s}}%
}\delta f_{s}.
\end{equation}%
It is useful to further elaborate this expression in view of the following
calculations, by pointing out the dependences on the particle velocity
components represented in cylindrical coordinates as $\mathbf{u}=\left(
u_{R},u_{\varphi },u_{z}\right) $. Thus, by defining the configuration-space
quantities%
\begin{eqnarray}
A &\equiv &\left( \frac{M_{s}cR}{Z_{s}e}\right) ^{2}V_{\varphi }^{2}+2\frac{%
M_{s}cR}{Z_{s}e}\psi V_{\varphi }+\psi ^{2}, \\
B &\equiv &\left( \frac{M_{s}cR}{Z_{s}e}\right) ^{2}2V_{\varphi }+2\frac{%
M_{s}cR}{Z_{s}e}\psi , \\
C &\equiv &\left( \frac{M_{s}cR}{Z_{s}e}\right) ^{2},  \label{C}
\end{eqnarray}%
we can write the expression (\ref{df-a}) in the compact notation%
\begin{eqnarray}
\delta f_{s} &=&-\alpha _{s}A-\alpha _{s}Bu_{\varphi }-\alpha
_{s}Cu_{\varphi }^{2}  \notag \\
&\equiv &\delta f_{A}+\delta f_{B}+\delta f_{C}.
\end{eqnarray}%
We finally have%
\begin{eqnarray}
\delta \pi _{s} &=&\frac{n_{M}M_{s}}{\left( 2\pi /M_{s}\right)
^{3/2}T_{s}^{3/2}}\int du_{R}\int du_{z}\int du_{\varphi }  \notag \\
&&\left[ \mathbf{uu}e^{-\frac{M_{s}\left( \mathbf{u}\right) ^{2}}{2T_{s}}%
}\left( \delta f_{A}+\delta f_{B}+\delta f_{C}\right) \right]  \notag \\
&\equiv &\delta \pi _{A}+\delta \pi _{B}+\delta \pi _{C},
\end{eqnarray}%
where in the last line we have omitted for simplicity the use of species
subscript "$s$", without possibility of misunderstanding.

The first term $\delta \pi _{A}$ does not contain additional contributions
proportional to particle velocity, and is simply proportional to $\pi _{Ms}$%
, and therefore to $P_{s}$. Explicit calculation in fact gives%
\begin{equation}
\delta \pi _{A}=P_{s}\left( -\alpha _{s}A\right) \underline{\underline{%
\mathbf{I}}}.
\end{equation}%
This contribution is isotropic as the leading-order Maxwellian term. From
the physical point of view, it originates from perturbative corrections of
the density associated with the fluid rotation and non-vanishing magnetic
poloidal flux. The second term $\delta \pi _{B}$ carries odd contributions
in the particle velocity, and therefore it does not contribute to the
pressure tensor. We have identically%
\begin{equation}
\delta \pi _{B}=0.
\end{equation}%
The third contribution $\delta \pi _{C}$ is the most relevant one, as it is
still diagonal, but not isotropic. Hence, we must compute separately the
three diagonal entries. We notice that the $RR$ and $zz$ components are the
same and take the following expression:%
\begin{equation}
\delta \pi _{C\left( RR\right) }=\delta \pi _{C\left( zz\right) }=\left(
-\alpha _{s}C\right) n_{Ms}\frac{T_{s}^{2}}{M_{s}}.
\end{equation}%
Instead, the $\varphi \varphi $ terms gives%
\begin{equation}
\delta \pi _{C\left( \varphi \varphi \right) }=n_{Ms}T_{s}\left( -\alpha
_{s}C\right) 3\frac{T_{s}}{M_{s}}.
\end{equation}%
Expressing the results in cylindrical coordinates $\left( R,\varphi
,z\right) $ we get%
\begin{equation}
\delta \pi _{C}=n_{Ms}T_{s}\left( -\alpha _{s}C\right) \frac{T_{s}}{M_{s}}%
\left( 
\begin{array}{ccc}
1 & 0 & 0 \\ 
0 & 3 & 0 \\ 
0 & 0 & 1%
\end{array}%
\right) .
\end{equation}%
The perturbative tensor $\delta \pi _{s}$ corresponding to case A is then
given by%
\begin{eqnarray}
\delta \pi _{s} &=&\delta \pi _{A}+\delta \pi _{C}  \notag \\
&=&n_{Ms}T_{s}\left( -\alpha _{s}A\right) \left( 
\begin{array}{ccc}
1 & 0 & 0 \\ 
0 & 1 & 0 \\ 
0 & 0 & 1%
\end{array}%
\right)  \notag \\
&&+n_{Ms}T_{s}\left( -\alpha _{s}C\right) \frac{T_{s}}{M_{s}}\left( 
\begin{array}{ccc}
1 & 0 & 0 \\ 
0 & 3 & 0 \\ 
0 & 0 & 1%
\end{array}%
\right) .
\end{eqnarray}

We can finally collect the leading-order and perturbative terms and obtain
the final tensor representation:%
\begin{eqnarray}
\underline{\underline{\mathbf{\Pi }}}_{s} &=&\pi _{Ms}+\delta \pi _{s} 
\notag \\
&=&n_{Ms}T_{s}\left[ 1-\alpha _{s}A-\alpha _{s}\frac{C}{M_{s}}T_{s}\right] 
\underline{\underline{\mathbf{I}}}  \notag \\
&&-2\alpha _{s}\frac{C}{M_{s}}n_{Ms}T_{s}^{2}\mathbf{e}_{\varphi }\mathbf{e}%
_{\varphi },  \label{ptensor-A}
\end{eqnarray}%
from which the non-isotropic character is evident. We notice that:

1)\ The pressure tensor is affected by the non-ideal contribution in two
different ways. The first one is an isotropic modification of the pressure
tensor, the second one is a non-isotropic contribution to the pressure
tensor.

2)\ The non-isotropic contribution is proportional to $C$, and therefore it
arises because of the angular momentum contribution to the canonical
momentum conservation. The isotropic corrections instead are proportional to 
$C$ and $A$, so that in the second case they arise due to the fluid angular
frequency and the presence of poloidal magnetic field.

3)\ The new non-isotropic corrections to the Maxwellian pressure exhibit
characteristic dependencies on the temperature and also configuration-space
dependences contained in the coefficients $A$ and $C$.

4) The occurrence of the anisotropy makes the plasma colder with respect to
the pure isotropic Maxwellian case, in the sense that there arises a
temperature depletion which is greater in the azimuthal direction (see also
discussion below on this point).

\subsection{Pressure anisotropy: case B}

In the case B the representation of $\delta f_{s}$ to be used in Eq.(\ref%
{deltap-0}) is provided by Eq.(\ref{df-b}). Again, in order to evaluate the
integrals explicitly we introduce the change of variables letting $\mathbf{%
u=v}-\mathbf{V}_{s}$, where $d\mathbf{u=}d\mathbf{v}$, so that expressing
the Maxwellian KDF we can write the formal expression%
\begin{equation}
\delta \pi _{s}=\frac{n_{M}M_{s}}{\left( 2\pi /M_{s}\right) ^{3/2}T_{s}^{3/2}%
}\int d^{3}u\mathbf{uu}e^{-\frac{M_{s}\left( \mathbf{u}\right) ^{2}}{2T_{s}}%
}\delta f_{s}.
\end{equation}%
It is useful to further elaborate this expression by pointing out the
dependences on the particle velocity components represented in cylindrical
coordinates as $\mathbf{u}=\left( u_{R},u_{\varphi },u_{z}\right) $. Thus,
by defining the configuration-space quantities%
\begin{eqnarray}
D &\equiv &\frac{M_{s}}{2B}V_{\varphi }^{2}, \\
E &\equiv &\frac{M_{s}}{2B}2V_{\varphi }, \\
F &\equiv &\frac{M_{s}}{2B},  \label{F}
\end{eqnarray}%
and noting that to current order $V_{R}=0$, after change of variables we can
write $\delta f_{s}$ in Eq.(\ref{df-b}) as%
\begin{eqnarray}
\delta f_{s} &=&-\alpha _{s}D-\alpha _{s}Eu_{\varphi }-\alpha
_{s}Fu_{\varphi }^{2}-\alpha _{s}Fu_{R}^{2}  \notag \\
&=&\delta f_{D}+\delta f_{E}+\delta f_{F\varphi }+\delta f_{F_{R}},
\end{eqnarray}%
where the meaning of notation is again understood. We finally have%
\begin{eqnarray}
\delta \pi _{s} &=&\frac{n_{Ms}M_{s}}{\left( 2\pi /M_{s}\right)
^{3/2}T_{s}^{3/2}}\int du_{R}\int du_{z}\int du_{\varphi }  \notag \\
&&\left[ \mathbf{uu}e^{-\frac{M_{s}\left( \mathbf{u}\right) ^{2}}{2T_{s}}%
}\left( \delta f_{D}+\delta f_{E}+\delta f_{F\varphi }+\delta
f_{F_{R}}\right) \right]  \notag \\
&\equiv &\delta \pi _{D}+\delta \pi _{E}+\delta \pi _{F\varphi }+\delta \pi
_{F_{R}}.
\end{eqnarray}

The first term $\delta \pi _{D}$ does not contain additional contributions
proportional to particle velocity, so that it is proportional to $\pi _{Ms}$%
, and therefore to $P_{s}$. This perturbative tensor is therefore isotropic
as the leading-order Maxwellian term. Explicit calculation in fact gives%
\begin{equation}
\delta \pi _{D}=n_{Ms}T_{s}\left( -\alpha _{s}D\right) \underline{\underline{%
\mathbf{I}}}.
\end{equation}%
The second term $\delta \pi _{E}$ carries odd powers in the particle
velocity, and therefore it does not contribute to the pressure tensor:%
\begin{equation}
\delta \pi _{E}=0.
\end{equation}%
The third contribution $\delta \pi _{F\varphi }$ is diagonal, but not
isotropic, requiring the separate calculation of the three diagonal terms.
The $RR$ and $zz$ entries are the same and yield%
\begin{equation}
\delta \pi _{F_{\varphi }\left( RR\right) }=\delta \pi _{F_{\varphi }\left(
zz\right) }=\left( -\alpha _{s}F\right) n_{Ms}\frac{T_{s}^{2}}{M_{s}}.
\end{equation}%
Instead, the $\varphi \varphi $ term gives%
\begin{equation}
\delta \pi _{F_{\varphi }\left( \varphi \varphi \right) }=n_{Ms}T_{s}\left(
-\alpha _{s}F\right) 3\frac{T_{s}}{M_{s}}.
\end{equation}%
Collecting the three results we get the matrix representation as follows:%
\begin{equation}
\delta \pi _{F_{\varphi }}=n_{Ms}T_{s}\left( -\alpha _{s}F\right) \frac{T_{s}%
}{M_{s}}\left( 
\begin{array}{ccc}
1 & 0 & 0 \\ 
0 & 3 & 0 \\ 
0 & 0 & 1%
\end{array}%
\right) .
\end{equation}%
Finally, the fourth contribution $\delta \pi _{F_{R}}$ is analogous to $%
\delta \pi _{F\varphi }$, but with entries exchanged. It therefore gives:%
\begin{equation}
\delta \pi _{F_{R}}=n_{Ms}T_{s}\left( -\alpha _{s}F\right) \frac{T_{s}}{M_{s}%
}\left( 
\begin{array}{ccc}
3 & 0 & 0 \\ 
0 & 1 & 0 \\ 
0 & 0 & 1%
\end{array}%
\right) .
\end{equation}%
Collecting now the three non-vanishing contributions, we can express the
resulting perturbative tensor $\delta \pi _{s}$ carrying non-isotropic
pressure due to magnetic-moment conservation as follows:%
\begin{eqnarray}
\delta \pi _{s} &=&\delta \pi _{D}+\delta \pi _{F\varphi }+\delta \pi
_{F_{R}}  \notag \\
&=&n_{Ms}T_{s}\left( -\alpha _{s}D\right) \left( 
\begin{array}{ccc}
1 & 0 & 0 \\ 
0 & 1 & 0 \\ 
0 & 0 & 1%
\end{array}%
\right)  \notag \\
&&+n_{Ms}T_{s}\left( -\alpha _{s}F\right) \frac{T_{s}}{M_{s}}2\left( 
\begin{array}{ccc}
2 & 0 & 0 \\ 
0 & 2 & 0 \\ 
0 & 0 & 1%
\end{array}%
\right) .
\end{eqnarray}

In conclusion, the combined contributions of the leading-order and
perturbative pressure tensors can be written in compact matrix notation as
the following non-isotropic tensor:%
\begin{eqnarray}
\underline{\underline{\mathbf{\Pi }}}_{s} &=&\pi _{Ms}+\delta \pi _{s} 
\notag \\
&=&n_{Ms}T_{s}\left[ 1-\alpha _{s}D-\alpha _{s}\frac{F}{M_{s}}2T_{s}\right] 
\underline{\underline{\mathbf{I}}}  \notag \\
&&-2\alpha _{s}\frac{F}{M_{s}}n_{Ms}T_{s}^{2}\left[ \mathbf{e}_{R}\mathbf{e}%
_{R}+\mathbf{e}_{\varphi }\mathbf{e}_{\varphi }\right] .  \label{ptensor-B}
\end{eqnarray}%
We notice that:

1)\ The pressure tensor is affected by the non-ideal contribution in two
different ways. The first one is an isotropic modification of the pressure
tensor, the second one is a non-isotropic contribution to the pressure
tensor.

2)\ The non-isotropic contribution is proportional to the
configuration-space function $F$, and therefore it arises because of the
magnetic moment conservation. The isotropic corrections instead are
proportional to $F$ and $D$, so that in the second case they arise due to
the fluid angular frequency.

3)\ The new non-isotropic corrections to the Maxwellian pressure exhibit
characteristic power-law dependences on the temperature.

4) As for the previous case A, also the magnetic-moment anisotropy
contributes to make the plasma colder with respect to the pure isotropic
Maxwellian case, with the depletion being non-isotropic and greater in the
radial and azimuthal directions.

\section{Polytropic representation of the EoS}

The analytical solutions found in previous section for the non-isotropic
pressure tensors of cases A and B can be regarded as equation of states for
the corresponding physical scenarios. For this reason, the same matrix
expressions provide also the sought equilibrium and species-dependent
kinetic closure condition for the fluid momentum equation. In this section
we go a step further and propose an iterative scheme to be implemented in
order to reach a polytropic representation for the tensors $\underline{%
\underline{\Pi }}_{s}$ in both Eq.(\ref{ptensor-A}) and Eq.(\ref{ptensor-B}%
). The goal is to represent each non-vanishing entry of the same tensors in
terms of assigned functions of mass density $\rho $ with a characteristic
power-law dependence, namely with a precise polytropic index $\Gamma $.

The starting point is the assumption of validity of polytropic EoS for the
Maxwellian isotropic pressure, namely the relation:%
\begin{equation}
P_{s}=n_{Ms}T_{s}\equiv \kappa \rho _{s}^{\Gamma }
\end{equation}%
where $\rho _{s}$ is the mass density, $\kappa $ is a suitable dimensional
function of proportionality and $\Gamma $ is the polytropic index. Notice
that both $\kappa $ and $\Gamma $ are in principle species-dependent
quantities, but nevertheless we omit the subscript "$s$" for convenience of
notation. From the previous equation we can then obtain a representation of
the temperature as follows%
\begin{equation}
T_{s}\equiv \kappa M_{s}\rho _{s}^{\Gamma -1},
\end{equation}%
where by definition $\rho _{s}=M_{s}n_{Ms}$, and hereon $\rho _{s}\equiv
\rho _{Ms}$ identifies the Maxwellian mass density function corresponding to 
$n_{Ms}$. We now introduce the assumption that we can iterate the same
polytropic relationship for the temperature also in the perturbative
non-isotropic corrections to the pressure tensor. In this way the polytropic
representation of the pressure tensor is reached.

Let us consider first the case A of tangential pressure anisotropy given by
Eq.(\ref{ptensor-A}). The result is as follows:%
\begin{eqnarray}
\underline{\underline{\mathbf{\Pi }}}_{s} &=&\kappa \rho _{s}^{\Gamma }\left[
1-\alpha _{s}A\right] \underline{\underline{\mathbf{I}}}-\alpha _{s}C\kappa
^{2}\rho _{s}^{2\Gamma -1}\underline{\underline{\mathbf{I}}}  \notag \\
&&-2\alpha _{s}C\kappa ^{2}\rho _{s}^{2\Gamma -1}\mathbf{e}_{\varphi }%
\mathbf{e}_{\varphi }.  \label{polyA-1}
\end{eqnarray}%
We see that the perturbative isotropic corrections modify the isotropic
polytropic relation with both the factor $-\alpha _{s}A$ and the new term $%
-\alpha _{s}C\kappa ^{2}\rho ^{2\Gamma -1}$ entering with the power-law of
the density as $\rho _{s}^{2\Gamma -1}$. The non-isotropic $\mathbf{e}%
_{\varphi }\mathbf{e}_{\varphi }-$term carries similarly a power of density
as $\rho _{s}^{2\Gamma -1}$.

It is also instructive to represent Eq.(\ref{polyA-1}) as%
\begin{equation}
\underline{\underline{\mathbf{\Pi }}}_{s}=p_{s}\underline{\underline{\mathbf{%
I}}}-2\alpha _{s}C\kappa ^{2}\rho _{s}^{2\Gamma -1}\mathbf{e}_{\varphi }%
\mathbf{e}_{\varphi },  \label{pi-finaleA}
\end{equation}%
where we define the isotropic pressure $p_{s}$ with perturbative corrections
as%
\begin{equation}
p_{s}\equiv P_{s}\left[ 1-\alpha _{s}A\right] -\alpha _{s}C\kappa ^{2}\rho
_{s}^{2\Gamma -1}.
\end{equation}%
This expression is instrumental for the evaluation of the contribution of $%
\underline{\underline{\mathbf{\Pi }}}_{s}$ to the equilibrium Euler momentum
equation.\ It provides a physically-based kinetic closure condition which
goes beyond the ideal-fluid case, by including in a consistent way the
non-isotropic perturbative corrections motivated by the kinetic solution. In
fact, the Euler equation contains the divergence of the pressure tensor,
namely $\nabla \cdot \underline{\underline{\mathbf{\Pi }}}_{s}$. In the
Maxwellian case this reduces simply to $\nabla P_{s}$. Instead, thanks to
Eq.(\ref{pi-finaleA}), now we have that%
\begin{equation}
\nabla \cdot \underline{\underline{\mathbf{\Pi }}}_{s}=\nabla p_{s}+\nabla
\cdot \left( -2\alpha _{s}C\kappa ^{2}\rho _{s}^{2\Gamma -1}\mathbf{e}%
_{\varphi }\mathbf{e}_{\varphi }\right) .
\end{equation}%
Under assumption of axisymmetric geometry and cylindrical coordinates $%
\left( R,\varphi ,z\right) $, then%
\begin{equation}
\nabla p_{s}=\left( \frac{\partial }{\partial R}p_{s},0,\frac{\partial }{%
\partial z}p_{s}\right) ,
\end{equation}%
while the divergence of the $\mathbf{e}_{\varphi }\mathbf{e}_{\varphi }$
gives%
\begin{equation}
\nabla \cdot \left( 2\alpha _{s}C\kappa ^{2}\rho _{s}^{2\Gamma -1}\mathbf{e}%
_{\varphi }\mathbf{e}_{\varphi }\right) =\left( \frac{2\alpha _{s}C\kappa
^{2}\rho _{s}^{2\Gamma -1}}{R},0,0\right) ,
\end{equation}%
and therefore%
\begin{equation}
\nabla \cdot \underline{\underline{\mathbf{\Pi }}}_{s}=\left( \frac{\partial 
}{\partial R}p_{s}+\frac{2\alpha _{s}C\kappa ^{2}\rho _{s}^{2\Gamma -1}}{R}%
,0,\frac{\partial }{\partial z}p_{s}\right) .
\end{equation}%
From this expression it is clear that the tangential pressure anisotropy
operating in the toroidal direction (the direction of symmetry)\ ultimately
can affect in a non-trivial way the non-ideal fluid dynamics with different
force terms transferred in the vertical and radial directions.

Let us now consider the case B of magnetic-moment generated temperature
anisotropy. After iteration of the polytropic representation we have%
\begin{eqnarray}
\underline{\underline{\mathbf{\Pi }}}_{s} &=&\kappa \rho _{s}^{\Gamma }\left[
1-\alpha _{s}D\right] \underline{\underline{\mathbf{I}}}-\alpha _{s}2F\kappa
^{2}\rho _{s}^{2\Gamma -1}\underline{\underline{\mathbf{I}}}  \notag \\
&&-2\alpha _{s}F\kappa ^{2}\rho _{s}^{2\Gamma -1}\left[ \mathbf{e}_{R}%
\mathbf{e}_{R}+\mathbf{e}_{\varphi }\mathbf{e}_{\varphi }\right] ,
\label{polyB-1}
\end{eqnarray}%
which represents the sought polytropic relation. Again, the perturbative
isotropic corrections modify the isotropic polytropic relation with the
factor $-\alpha _{s}D$ and the new term $-\alpha _{s}2F\kappa ^{2}\rho
_{s}^{2\Gamma -1}$ depending on the mass density as $\rho _{s}^{2\Gamma -1}$%
. The non-isotropic $\mathbf{e}_{\varphi }\mathbf{e}_{\varphi }$ and $%
\mathbf{e}_{R}\mathbf{e}_{R}-$terms carry a power of density as $\rho
_{s}^{2\Gamma -1}$. By defining now the isotropic pressure $p_{s}$ with
perturbative corrections as%
\begin{equation}
p_{s}\equiv P_{s}\left[ 1-\alpha _{s}D\right] -\alpha _{s}F\kappa ^{2}\rho
_{s}^{2\Gamma -1},
\end{equation}%
the pressure tensor can also be written as%
\begin{equation}
\underline{\underline{\mathbf{\Pi }}}_{s}=p_{s}\underline{\underline{\mathbf{%
I}}}-2\alpha _{s}F\kappa ^{2}\rho _{s}^{2\Gamma -1}\left[ \mathbf{e}_{R}%
\mathbf{e}_{R}+\mathbf{e}_{\varphi }\mathbf{e}_{\varphi }\right] .
\label{pi-finaleB}
\end{equation}%
Its role in the Euler momentum equation can be readily displayed by
evaluating the divergence $\nabla \cdot \underline{\underline{\mathbf{\Pi }}}%
_{s}$ in cylindrical coordinates. We obtain%
\begin{equation}
\nabla \cdot \underline{\underline{\mathbf{\Pi }}}_{s}=\nabla p_{s}+\nabla
\cdot \left( -2\alpha _{s}F\kappa ^{2}\rho _{s}^{2\Gamma -1}\left[ \mathbf{e}%
_{R}\mathbf{e}_{R}+\mathbf{e}_{\varphi }\mathbf{e}_{\varphi }\right] \right)
.
\end{equation}%
As usual, assuming axisymmetry, the isotropic contribution simply gives%
\begin{equation}
\nabla p_{s}=\left( \frac{\partial }{\partial R}p_{s},0,\frac{\partial }{%
\partial z}p_{s}\right) .
\end{equation}%
Instead, the divergence of the remaining tensor components gives%
\begin{eqnarray}
&&\left. \nabla \cdot \left( -2\alpha _{s}F\kappa ^{2}\rho _{s}^{2\Gamma -1} 
\left[ \mathbf{e}_{R}\mathbf{e}_{R}+\mathbf{e}_{\varphi }\mathbf{e}_{\varphi
}\right] \right) =\right. \\
&&\left. \left( \frac{2\alpha _{s}F\kappa ^{2}\rho _{s}^{2\Gamma -1}}{R}-%
\frac{1}{R}\frac{\partial }{\partial R}\left( 2\alpha _{s}F\kappa ^{2}\rho
_{s}^{2\Gamma -1}R\right) ,0,0\right) ,\right.  \notag
\end{eqnarray}%
and therefore%
\begin{equation}
\nabla \cdot \underline{\underline{\mathbf{\Pi }}}_{s}=\left( \binom{\frac{%
\partial }{\partial R}p_{s}+\frac{2\alpha _{s}F\kappa ^{2}\rho _{s}^{2\Gamma
-1}}{R}}{-\frac{1}{R}\frac{\partial }{\partial R}\left( 2\alpha _{s}F\kappa
^{2}\rho _{s}^{2\Gamma -1}R\right) },0,\frac{\partial }{\partial z}%
p_{s}\right) .
\end{equation}%
From this expression it is clear that the differential contribution of the
anisotropy in the radial and toroidal directions, generated by
magnetic-moment conservation in a purely-vertical magnetic field, is
ultimately translated in the vertical and radial directions, with different
force terms affecting the configuration of the non-ideal fluid system.

\section{Astrophysical implications}

In this section we consider qualitative analysis of the main astrophysical
implications of the theory developed here and the polytropic representation
of the EoS. The discussion applies to astrophysical magnetized plasmas
characterized by azimuthal differential rotation and belonging to
accretion-disc systems or hot-corona environments surrounding compact
objects. For them, the solutions obtained here can provide a useful
background for the prediction/interpretation of non-ideal plasma properties.
These systems in fact are expected to give rise to collisionless equilibrium
states that can approach a nearly-Maxwellian configuration on the
characteristic collision time scales. In such a case a plasma system can
still develop phase-space anisotropies characteristic of collisionless
states, but being at the same time sufficiently close to the Maxwellian
distribution, in such a way that non-ideal deviations can be regarded as
weak in an appropriate ordering scheme. When these occurrences take place,
the physical assumptions underlying the present theoretical model apply. The
goal is to prove that non-ideal effects of the type pointed out here,
although small in an appropriate sense, can nevertheless introduce relevant
implications beyond the simple ideal-fluid assumption.

\subsection{Estimate of plasma temperature}

The first application concerns the problem of the observational estimate of
temperature profiles of magnetized plasma systems. Observational data of hot
plasma systems are usually inferred from x-ray observations and are
associated to bremsstrahlung radiative mechanisms. Let us denote the real
system temperature as $T_{obs}$. From the analysis carried out above we know
that in general, even for weak deviations from the Maxwellian state, we can
write a relationship of the type%
\begin{equation}
T_{obs}=T_{M}-\left\vert \delta T\right\vert ,
\end{equation}%
where $T_{M}$ is the purely Maxwellian temperature and $\delta T$ is the
correction due to anisotropy in non-ideal fluids. The previous relation is
purely symbolic at this stage, although it could in principle apply to each
temperature component of the pressure tensor. We notice a characteristic
feature that arises from the previous analysis, namely that the contribution
of the non-ideal fluid temperature is always negative with respect to the
Maxwellian case. This justifies the use of absolute value and the minus sign
singled out in front of it. Thus, the action of a phase-space anisotropy is
that of reducing the Maxwellian temperature, causing a depletion in the
observed temperature. From the mathematical point of view this is a
consequence of the integrability condition (i.e., convergence on velocity
space)\ imposed on the equilibrium Gaussian-like KDF $f_{\ast s}$ for the
same existence of fluid moments. Instead, from the physical point of view
this can be explained by recalling that in kinetic theory the temperature
has a statistical meaning as a velocity dispersion. When additional kinetic
or phase-space constraints are present and retained in the distribution
function, then the collective dynamics of single charges is more constrained
and there remains less freedom to generate velocity dispersion around the
mean value with respect to the unconstrained Maxwellian solution. However,
we can see that if one interpolates the observational data assuming a purely
Maxwellian solution and neglecting the non-isotropic contributions, then the
estimate of energy content of a system is wrongly increased. In fact we have
that $T_{M}=T_{obs}+\left\vert \delta T\right\vert $. The error depends on
how big is the non-ideal fluid contribution with respect to the Maxwellian
one. Under validity of the present theory, it would depend on the magnitude
of the parameter $\varepsilon \ll 1$, up to an acceptable value of $%
\varepsilon \sim 0.1$.

\subsection{Distinction between sources of anisotropy}

The second application concerns the analysis of the physical implication of
the polytropic representation of the EoS in cases A and B for the
corresponding sources of pressure anisotropy. The issue belongs to the more
general problem of relating the polytropic index of the EoS with temperature
and pressure anisotropies arising in astrophysical plasmas \cite{Liva2}. In
fact, comparing the solutions (\ref{polyA-1}) and (\ref{polyB-1}) we notice
that the perturbative non-isotropic corrections $\delta \pi $ enter the
pressure in both cases with the same power-law dependence on the fluid mass
density and polytropic index, equal to $2\Gamma -1$. Namely, for the
polytropic non-isotropic contribution we obtain the general symbolic
expression as far as the density dependence is concerned:%
\begin{equation}
\delta \pi \sim \rho ^{2\Gamma -1}.
\end{equation}%
This means that there is a sort of degeneracy among these kinds of
non-isotropic pressure effects. The physical explanation lies in the fact
that the phase-space anisotropies $\delta f_{s}$ in the respective cases
depend always on the square of particle-velocity components. Ultimately,
this kind of dependence implies a term of non-isotropic pressure
proportional to $\rho _{s}^{2\Gamma -1}$. It means that, under the
conditions of validity of the present theoretical model, in principle it is
not possible to distinguish one effect of temperature anisotropy from the
other from observations of temperature profiles over density profiles.
Hence, it is not possible to exclude or select an onset non-ideal fluid
temperature anisotropy effect only on the basis of the polytropic density
dependence.

This kind of degeneracy however can be resolved by combining other
fundamental features. Namely, the distinction of directional temperatures
(i.e., the directions of anisotropies)\ and their spatial profiles with
respect to the configuration-space functions ($A,B,C$)\ and ($D,E,F$)
respectively in cases A and B, as well as the resulting dynamical
contribution to fluid forces in the Euler momentum equation. It is
instructive to illustrate in more detail the argument. Let us consider for
this purpose the two representations (\ref{pi-finaleA}) and (\ref{pi-finaleB}%
). We restrict the analysis to the non-isotropic contributions entering the $%
\mathbf{e}_{\varphi }\mathbf{e}_{\varphi }$ component of the pressure
tensor, which are denoted for simplicity as $\delta \pi _{A}$\ and $\delta
\pi _{B}$\ respectively. In this example case we can additionally assume,
without loss of generality, that in both expressions the dimensional
coefficients $\alpha _{s}$\ and $\kappa $\ are constant, so that we can set
identically $\alpha _{s}=const.$\ and $\kappa =const.$\ Under these
assumptions the non-isotropic $\mathbf{e}_{\varphi }\mathbf{e}_{\varphi }$%
-components scale as%
\begin{eqnarray}
\text{Case A}\text{:\ \ \ } &&\delta \pi _{A}\sim C\rho ^{2\Gamma -1}, \\
\text{Case B}\text{: \ \ } &&\delta \pi _{B}\sim F\rho ^{2\Gamma -1}.
\end{eqnarray}

We now invoke the definitions of the configuration-space functions $C$\ and $%
F$. From Eq.(\ref{C}) we see that $C$\ scales with the cylindrical radial
coordinate as $C\sim R^{2}$. A different kind of dependence instead
characterizes the function $F$, which from Eq.(\ref{F}) depends on the
magnitude of the magnetic field $B$\ as $F\sim B^{-1}$. We then implement
the solution of the magnetic field reported in Eq.(\ref{bext-2}) and assume
that the dominant spatial dependence of $B$\ is on the radial coordinate,
namely $B=B_{z}\equiv B_{z}\left( R,\varepsilon ^{k}z\right) $, with $k\geq
1 $. If we further assume a power-law radial dependence with power exponent $%
\beta $, we have that $B\sim R^{-\beta }$. This implies that necessarily $%
F\sim R^{\beta }$. Hence, we can finally compare the explicit functional
form of the pressure contributions in the two cases by writing:%
\begin{eqnarray}
\text{Case A}\text{:\ \ \ } &&\delta \pi _{A}\sim R^{2}\rho ^{2\Gamma -1}, \\
\text{Case B}\text{: \ \ } &&\delta \pi _{B}\sim R^{\beta }\rho ^{2\Gamma
-1}.
\end{eqnarray}

These expressions prove that, ultimately, the spatial dependence of the
functions $C$\ and $F$\ can effectively produce a concrete distinction
between the two sources of anisotropy, although both exhibit the same
power-law polytropic dependence on mass density. In particular, we notice
that in Case A the radial dependence is uniquely fixed by $C\sim R^{2}$.
This follows by the fact that the same function $C$\ is associated with the
conservation of particle canonical momentum in the axisymmetric torus.
Instead, in Case B the power-law is associated with the behavior of the
magnetic field by $F\sim R^{\beta }$, where the exponent $\beta $\ is still
to be assigned. In the anisotropy associated with magnetic-moment
conservation there arises therefore more freedom for the resulting spatial
variation of pressure anisotropy. Finally, as a concluding illustration,
consider for example the set in which $\Gamma =2$, $\beta =3$\ and $\rho
\sim R^{-2}$. Then, in both cases the leading-order polytropic pressure
would scale as $P\sim R^{-4}$. The first-order non-isotropic corrections
would be instead such that%
\begin{eqnarray}
\text{Case A}\text{:\ \ \ } &&\delta \pi _{A}\sim R^{-4}, \\
\text{Case B}\text{: \ \ } &&\delta \pi _{B}\sim R^{-3}.
\end{eqnarray}%
As a consequence, the anisotropy in Case B would scale more slowly in radial
direction than the leading-order pressure or Case A. This ultimately
reflects the different physical origin of the two non-isotropic mechanisms
discussed here. It is therefore expected that similar conclusions should
apply in concrete physical/astrophysical scenarios, where the functional
form of pressure anisotropies can be evaluated analytically or numerically
for sets of fluid parameters expressed by realistic values of polytropic
index and profiles of mass density and magnetic field. This task will be the
subject of forthcoming studies dedicated to a better understanding of the
physical significance of pressure anisotropy in non-ideal fluids.

\section{Conclusions}

In this paper the problem of reaching a polytropic representation for the
equation of state (EoS) of non-ideal fluids has been addressed, whereby the
pressure is expressed as a power-law function of the fluid mass density with
assigned polytropic index. The case of non-relativistic collisionless
plasmas belonging to axisymmetric toroidal structures and subject to
electromagnetic and gravitational fields has been considered. A statistical
treatment based on kinetic Vlasov theory for the determination of
equilibrium solutions for the species kinetic distribution function (KDF)
has been implemented. The physical mechanism responsible for the onset of
non-ideal fluid properties is identified with the occurrence of microscopic
phase-space anisotropies. The latter ones constraint the functional form of
the equilibrium KDF to be different from a Maxwellian distribution. As a
consequence, it is shown that the resulting continuum fluid solution becomes
characterized by a non-isotropic pressure tensor, in contrast to the
customary isotropic scalar pressure that arises in purely Maxwellian
configurations. It must be recalled that the pressure tensor represents the
closure condition that is needed for the integration of continuity and
momentum fluid equations, and therefore for the understanding of density and
velocity fluid profiles. Therefore, it is understood that its knowledge
based on kinetic approach provides insights into the physical content of the
system as far as microscopic single-particle and collective plasma dynamics
is concerned.

Two different source mechanisms of pressure anisotropy have been identified
and compared. They are associated respectively with the conservations of
particle canonical momentum due to axial symmetry assumption and magnetic
moment due to Larmor rotation of charges around magnetic field lines.
Suitable ordering assumptions have been discussed which allow for the
construction of analytical solutions for the corresponding KDFs by
implementation of a Chapman-Enskog expansion of the exact equilibrium
solution around a leading-order Maxwellian distribution. The technique has
the advantage of permitting to single out the role of phase-space
anisotropies, namely the non-ideal features of the kinetic distribution, and
to display their physical meaning and statistical role. In this way, also
corresponding analytical solutions for the pressure tensors of the two
configurations have been constructed. As a remarkable outcome, it has been
proved that, despite the tensorial character of the pressure, it is still
possible to represent the EoS in polytropic form, namely as an assigned
function of density power-law.

Finally, qualitative astrophysical implications of these outcomes has been
discussed. The first issue is about the inference of temperature and
pressure profiles in real systems, e.g., based on astronomical observational
data, with respect to interpolation based on the assumption of
purely-Maxwellian states. The second issue instead is about the possibility
of distinguishing pressure anisotropies generated by different non-ideal
effects on the basis of knowledge of mass-density profiles.

The results proposed here have been established on a comprehensive
theoretical background and they are expected to have a wide physical
relevance in plasma physics, fluid dynamics and astrophysical plasmas. For
this reason, the present kinetic theory is susceptible of further
investigation and is expected to help gaining insights into the complex
dynamics governing magnetized plasmas and fluids, with particular reference
to astrophysical scenarios. In addition, the analytical determination of
non-isotropic pressure tensors based on kinetic approach can also provide a
convenient framework to hydrodynamics and magneto-hydrodynamics studies of
non-ideal astrophysical fluids, e.g., through the prescription of
corresponding equations of state and fluid closure conditions. In fact, this
prescription can effectively become relevant in numerical or semi-analytical
studies of fluid dynamics which demand a polytropic representation for the
equation of state to warrant explicit integration of the differential
equations solving for the fluid density and velocity field profiles. In
addition, at the same time these kind of studies can gain physical relevance
by implementing a precise expression for the pressure tensor that goes
beyond the ideal-fluid solution with the inclusion of well-defined non-ideal
fluid effects.

\bigskip

\textbf{AUTHOR'S CONTRIBUTIONS}

All authors contributed equally to this work.

The authors have no conflicts to disclose.

\textbf{ACKNOWLEDGMENTS}

CC, JK and ZS acknowledge the support of the Research Centre for Theoretical
Physics and Astrophysics, Institute of Physics, Silesian University in
Opava, Czech Republic.

\textbf{DATA AVAILABILITY}

The data that support the findings of this study are available within the
article.

\end{document}